\documentclass[%
reprint,
nofootinbib,
mathtools,
amsmath,
amssymb,
prd,
longbibliography,
superscriptaddress]{revtex4-2}

\usepackage{booktabs}
\usepackage{mathrsfs}
\usepackage{graphicx}
\usepackage{dcolumn}
\usepackage{bm}
\usepackage{lipsum}
\usepackage{graphics}
\usepackage{bbm}
\usepackage[dvipsnames]{xcolor}
\usepackage{color}
\usepackage{microtype}
\usepackage{IEEEtrantools}
\usepackage{mathrsfs}
\usepackage{amsthm}
\usepackage{enumitem}
\usepackage[normalem]{ulem}
\usepackage[pdftex,breaklinks,colorlinks,
linkcolor=Blue,
citecolor=teal,
anchorcolor=red,
urlcolor=cyan]{hyperref}
\usepackage{orcidlink}

\begin{document}

\title{Spin-aligned inspiral waveforms from self-force and post-Newtonian theory}
\author{Loïc Honet\,\orcidlink{0009-0007-2863-6085}}
\affiliation{%
 Universit\'e Libre de Bruxelles, BLU-ULB Brussels Laboratory of the Universe, C.P. 231, B-1050 Bruxelles, Belgium
}%
\author{Josh Mathews\,\orcidlink{0000-0002-5477-8470}}
\affiliation{Department of Physics, National University of Singapore, 21 Lower Kent Ridge Rd, Singapore 119077}

\author{Geoffrey Comp\`ere\,\orcidlink{0000-0002-1977-3295}}
\affiliation{%
 Universit\'e Libre de Bruxelles, BLU-ULB Brussels Laboratory of the Universe, C.P. 231, B-1050 Bruxelles, Belgium
}
\author{Adam Pound\,\orcidlink{0000-0001-9446-0638}}
\affiliation{School of Mathematical Sciences and STAG Research Centre, University of Southampton, Southampton, United Kingdom, SO17 1BJ}
\author{Barry Wardell\,\orcidlink{0000-0001-6176-9006}}
\affiliation{School of Mathematics \& Statistics, University College Dublin, Belfield, Dublin 4, Ireland, D04 V1W8}

\author{Gabriel Andres Piovano\,\orcidlink{0000-0003-1782-6813}}
\affiliation{%
 Universit\'e Libre de Bruxelles, BLU-ULB Brussels Laboratory of the Universe, C.P. 231, B-1050 Bruxelles, Belgium
}%
\affiliation{
Physique de l’Univers, Champs et Gravitation, Universit\'{e} de Mons – UMONS,
Place du Parc 20, 7000 Mons, Belgium
}%
\author{Maarten van de Meent\,\orcidlink{0000-0002-0242-2464}}
\affiliation{Center of Gravity, Niels Bohr Institute, Blegdamsvej 17, 2100 Copenhagen, Denmark}
\affiliation{Max Planck Institute for Gravitational Physics (Albert Einstein Institute), Am M\"uhlenberg 1, Potsdam 14476, Germany}
\author{Niels Warburton\,\orcidlink{0000-0003-0914-8645}}
\affiliation{School of Mathematics \& Statistics, University College Dublin, Belfield, Dublin 4, Ireland, D04 V1W8}
\date{\today}

\begin{abstract}
   We present the state-of-the-art waveform model \texttt{WaSABI-C} for quasicircular inspirals of spinning black hole binaries with aligned or anti-aligned spins. 
   Our model synthesizes the most up-to-date first- and second-order gravitational self-force results with high-order post-Newtonian expansions through a systematic hybridization procedure. This approach captures both strong-field and weak-field dynamics with high fidelity, enabling accurate modeling of spin-(anti)aligned inspirals across a wide parameter space. The resulting waveforms mark a significant advance in the precision of self-force-based templates, providing critical input for the detection and interpretation of gravitational waves from compact binaries with future observatories such as LISA and ET. We accompany this work with the release of \href{https://bhptoolkit.org/WaSABI/}{\textsc{WaSABI}} (Waveform Simulations of Asymmetric Binary Inspirals), a public package implementing our model for community use and further development.
\end{abstract}

\maketitle

\emph{Introduction.}
The accurate modeling of gravitational waves from compact binary systems is essential for both the detection and interpretation of signals by current and future gravitational wave (GW) observatories. In particular, systems with extreme or intermediate mass ratios---such as stellar-mass objects orbiting massive black holes (BHs)---are key sources for upcoming missions like the Laser Interferometer Space Antenna (LISA)~\cite{LISA:2024hlh} and, in the case of intermediate mass ratios, for the Einstein Telescope (ET) \cite{Abac:2025saz} as well. These systems will be especially valuable probes of strong-field gravity, enabling precision tests of general relativity and the nature of BHs~\cite{LISA:2022kgy}, extending the first such landmark tests recently performed with the LVK detectors~\cite{KAGRA:2025oiz}. 

A critical challenge in this regime is the construction of waveform models that are both accurate and computationally efficient over the long inspiral phase, where the binary's orbital dynamics are governed by a combination of weak-field and strong-field gravitational physics~\cite{LISAConsortiumWaveformWorkingGroup:2023arg}. Even now, current LVK detectors have started observing coalescence events from systems with mass ratios as high as 1:27~\cite{LIGOScientific:2021djp}, a range in which only very limited model validation has been possible~\cite{Albertini:2022rfe,Islam:2022laz,Nagar:2022icd,Rink:2024swg}.

Gravitational self-force (SF) theory provides a rigorous framework for modeling systems with sufficiently small mass ratios~\cite{Poisson:2011nh,Barack:2018yvs,Pound:2021qin}, and it plays a central role in waveform modeling for extreme-mass-ratio inspirals (EMRIs). Recent advances have yielded linear-order SF results for spinning binaries~\cite{Piovano:2021iwv,Skoupy:2023lih,vandeMeent:2015lxa,vandeMeent:2017bcc} and quadratic-order SF results for nonspinning configurations~\cite{Pound:2019lzj,Warburton:2021kwk, Wardell:2021fyy}. However, SF waveform models are limited by lack of second-order SF results for spinning primary BHs, which will be required for accurate modelling of astrophysically realistic binaries~\cite{LISA:2022yao,LISAConsortiumWaveformWorkingGroup:2023arg}. Moreover, SF models lose accuracy for mass ratios near unity and for signals that include the early, weak-field portion of the inspiral~\cite{Albertini:2022rfe}.
Post-Newtonian (PN) theory, which is valid in the weak-field and comparable-mass regime~\cite{blanchet2024postnewtoniantheorygravitationalwaves}, complements the SF approach by providing analytically tractable expressions that can be hybridized with strong-field SF data to extend model validity across the parameter space.

In this Letter we present a fast, first-principles waveform model that targets the long inspiral phase of mass-asymmetric binaries while maintaining high fidelity to state-of-the-art waveforms of comparable-mass binaries. The model, \texttt{WaSABI-C} (Waveform Simulations of Asymmetric Binary Inspirals $-$ Circular), builds on companion papers~\cite{PaperII,PaperIII} and incorporates both first- and second-order SF results, hybridized with high-order PN expansions for quasicircular inspirals in which both black holes are spinning, with spins aligned or anti-aligned with the orbital angular momentum. This configuration is astrophysically relevant~\cite{LIGOScientific:2025pvj} and simplifies the modeling of spin effects while retaining key physical features. The resulting waveform model is accurate across a wide range of mass ratios, spins, and orbital separations, providing a powerful new tool for EMRI and intermediate-mass-ratio inspiral (IMRI) science. As a first-of-its-kind, \texttt{WaSABI-C} serves as a milestone towards a more comprehensive model including precession, eccentricity, and the merger-ringdown stage of the waveform~\cite{Compere:2021zfj, Kuchler:2024esj, Lhost:2024jmw,Kuchler:2025hwx,Roy:2025kra}.

To support the use and further development of our model, we introduce the public implementation \href{https://bhptoolkit.org/WaSABI/}{\textsc{WaSABI}}~\cite{BHPT_WaSABI} written in \textsc{Mathematica}. \textsc{WaSABI} provides a flexible and accessible platform for computing inspiral waveforms, enabling immediate integration into data-analysis-ready software platforms such as the FastEMRIWaveforms (FEW) package~\cite{Katz:2021yft, Chapman-Bird:2025xtd} and facilitating comparison with numerical relativity, effective-one-body, and other semi-analytical models. With this work, we advance the precision and accessibility of SF-based waveform modeling and lay groundwork for high-accuracy GW science in the era of third-generation detectors.

\emph{Formulation.} We use geometric units $G=c=1$. We consider a quasicircular, compact binary composed of two spinning black holes orbiting one another with a frequency $\Omega$. The primary, heavier BH has mass $m_1$ and dimensionless spin $\chi_1=S_1/(m_1)^2$; the secondary, lighter object has mass $m_2 \leq m_1$ and dimensionless spin $\chi_2=S_2/(m_2)^2$. We assume the spins are aligned or anti-aligned with the orbital angular momentum, which implies that the system is confined to a plane, with no spin precession. We also define the total mass $M=m_1+m_2$ and the symmetric mass ratio $\nu = m_1 m_2/M^2$. 

In both the PN ($M\Omega\ll 1$) and small-mass-ratio ($m_2/m_1\ll 1$) limits, the binary evolution and waveform have a multiscale form~\cite{Miller:2020bft,Pound:2021qin,blanchet2024postnewtoniantheorygravitationalwaves} governed by energy and angular-momentum flux-balance laws. Each BH's mass and spin evolve slowly due to absorption of gravitational radiation through its horizon, at the rates $dm_i/dt = {\cal F}_{{\cal H}_i}$ and $dS_i/dt =\mathcal G_{{\cal H}_i}$
~\cite{Teukolsky:1974yv,Thorne:1984mz,Tagoshi:1997jy,Alvi:2001mx,Poisson:2004cw, Ashtekar:2004cn},  where ${\cal F}_{{\cal H}_i}$ and ${\cal G}_{{\cal H}_i}$ are the energy and angular momentum fluxes through the $i$th BH's horizon, respectively; the quantities $d\chi_i/dt$ are then deduced by the chain rule. The slow evolution of the orbital frequency is determined from the binary's binding energy $E := M_{\rm B} - m_1 - m_2$, where $M_{\rm B}$ is the Bondi mass. Using $E=E(m_i,\chi_i,\Omega)$ and the Bondi-Sachs mass-loss formula $dM_{\rm B}/dt=-{\cal F}_\infty$~\cite{1962RSPSA.269...21B,1962RSPSA.270..103S}, where ${\cal F}_\infty$ is the GW energy flux to infinity, we apply the chain rule to $dE/dt$ and rearrange to find
\begin{equation}\label{Omega dot}
    \frac{d\Omega}{dt} = -\frac{{\cal F} + {\cal F}_{{\cal H}_i}\partial E/\partial m_i + \dot \chi_i \, \partial E/\partial \chi_i}{\partial E/\partial \Omega},
\end{equation}
with ${\cal F}\equiv{\cal F}_\infty + {\cal F}_{{\cal H}_1}+{\cal F}_{{\cal H}_2}$ and $i=1,2$ summed over. Here $t$ is a global time variable that penetrates the BHs' horizons and asymptotes to retarded time at future null infinity, where the waveform is extracted. The emitted GW strain, at a position $(r,\vartheta,\varphi)$ relative to the binary, then takes the form 
\begin{equation}\label{h}
h_+-i h_\times = \frac{1}{r}\sum_{\ell m}H_{\ell m}(m_i,\chi_i,\Omega)e^{-im\phi} {}_{-2}Y_{\ell m}(\vartheta,\varphi),
\end{equation}
where $\phi$ is the binary's orbital phase, related to $\Omega$ by $d\phi/dt=\Omega$. The complex amplitudes $H_{\ell m}$ slowly evolve with the system's parameters, and ${}_{-2}Y_{\ell m}$ are spin-weighted spherical harmonics. Each $\ell m$ mode is written as $h_{\ell m} = H_{\ell m}e^{-im\phi}$.

In line with the above generic structure, we formulate our waveform-generation scheme using the multiscale framework for spinning, asymmetric-mass binary inspirals developed in \cite{Pound:2021qin,Mathews:2021rod,Mathews:2025nyb}. This framework splits the problem into a computationally heavy ``offline" step followed by a computationally light ``online" step. For the quasicircular systems we consider, the offline step consists of solving the Einstein equations to obtain the fluxes $\mathcal F_\infty$, $\mathcal F_{\mathcal{H}_i}$, and $\mathcal G_{\mathcal{H}_i}$, the binding energy $E$, and the complex waveform amplitudes $H_{\ell m}$ as functions of the binary's slowly evolving mechanical parameters $(m_i,\chi_i,\Omega)$. That data is stored on a grid of parameter values, and the online step then consists of solving the differential equation~\eqref{Omega dot}, together with those for the masses and spins, to rapidly obtain the time dependence of the waveform~\eqref{h} (in a few tens of milliseconds~\cite{Katz:2021yft}).

Here, rather than working directly with $(m_i,\chi_i,\Omega)$, we adopt  $J_a \equiv (M, \nu, \chi_i,\omega)$ as our default set of parameters, where $\omega\approx M\Omega$ is one-half the (adimensionalized) frequency of the waveform's $(\ell,m)=(2,2)$ mode, defined as $\omega\equiv M d\psi/dt$ in terms of the $(2,2)$-mode phase $\psi\equiv-\frac{1}{2} {\rm arg}(h_{22})\approx \phi$. Our motivations for these choices are twofold: formulating SF expansions in terms of the total mass $M$ and symmetric mass ratio $\nu$ is known to dramatically improve accuracy of SF-based waveforms in the  comparable-mass regime~\cite{vandeMeent:2020xgc,Wardell:2021fyy}; and working with waveform phases and frequencies facilitates a more gauge-invariant hybridization between PN and SF information.

Evolution equations for $J_a(t)$ can be easily obtained from Eq.~\eqref{Omega dot} and the horizon-flux-balance laws. The set of ordinary differential equations to be solved in the online waveform-generation then take the form 
\begin{equation}\label{eq:ODE}
    \begin{split}
       \frac{d\psi}{dt} &=\omega/ M,\quad \frac{d\omega}{dt} =F_\omega(J_a),\\ 
        \frac{d\chi_i}{dt} &= F_{\chi_i}(J_a),\  
        \frac{d\nu}{dt} = F_\nu(J_a),         \ \frac{dM}{dt} = F_{M}(J_a), 
    \end{split}
\end{equation}
with initial conditions $\psi(t_0)=\psi_0$ and $J_a(t_0)=J_{0a}$. Similarly, we rewrite the waveform~\eqref{h} in terms of $\psi$ and $J_a$ as $h_{\ell m}(t)=A_{\ell m}(J_a(t))e^{-im\psi(t)}$. 

The forcing functions $F_a$ depend upon the binding energy $E$, the energy fluxes $\mathcal F_\infty$, $\mathcal F_{\mathcal H_i}$, and the angular momentum fluxes $\mathcal G_{\mathcal H_i}$, which we determine next, together with the amplitudes $A_{\ell m}$, from a hybridization of SF and PN data at all available orders in each expansion.

\emph{Hybridization.}
Our hybridization procedure is detailed in the companion paper~\cite{PaperIII} and the Supplementary Material. It involves two layers. On the top layer, we write the forcing functions $F_a$ in Eq.~\eqref{eq:ODE} in terms of $(E,{\cal F}_\infty,{\cal F}_{{\cal H}_i},{\cal G}_{{\cal H}_i})$ \emph{exactly}, without applying any SF or PN expansions. On the bottom layer, we build SF$+$PN composite expansions for $E$, ${\cal F}_\infty$, ${\cal F}_{{\cal H}_i}$, ${\cal G}_{{\cal H}_i}$, and the modulus and phase of $A_{\ell m}$. 

The composite expansions are constructed as follows. In our variables, an SF expansion is a series in $\nu$, while a PN expansion is a series in $x\equiv \omega^{2/3}$. For example, the binding energy and flux to infinity have the expansions
\begin{subequations}\label{E expansions}
\begin{align}
E &= \overbrace{M\nu E^{(0)}(x,\chi_1)}^{\text{0PA (0SF)}} + \overbrace{M\nu^2E^{(1)}(x,\chi_1,\chi_2)}^{\text{1PA (1SF)}}+\ldots\\
&=  -\underbrace{\frac{M}{2} \nu x}_{\text{0PN}} + \underbrace{\frac{M}{2}\left(\frac{3}{4}+\frac{\nu}{12}\right)\nu x^2}_{\text{1PN}}+\ldots
\end{align}
\end{subequations}
and
\begin{subequations}\label{Finfty expansions}
\begin{align}
{\cal F}_{\infty} &= \overbrace{\nu^2{\cal F}^{(0)}_{\infty}(x,\chi_1)}^{\text{0PA (1SF)}} + \overbrace{\nu^3{\cal F}^{(1)}_{\infty}(x,\chi_1,\chi_2)}^{\text{1PA (2SF)}}+\ldots\\[.33em]
&= \underbrace{\frac{32}{5}\nu^2x^5}_{\text{0PN}} -\underbrace{\left(\frac{2494}{105}+\frac{56 \nu }{3}\right)\nu^2 x^6}_{\text{1PN}} + \ldots
\end{align}
\end{subequations}
A composite expansion of each function is obtained by adding the SF and PN expansions together and subtracting doubly counted common terms, as is standard in asymptotic methods~\cite{KevorkianCole}. 

In Eqs.~\eqref{E expansions} and \eqref{Finfty expansions} we follow traditional nomenclature in SF theory~\cite{Hinderer:2008dm,LISAConsortiumWaveformWorkingGroup:2023arg}: An $n$th post-adiabatic ($n$PA) term contributes to $d\omega/dt$ at order $\nu^n$ \emph{relative} to the leading, adiabatic (0PA) dynamics, $d\omega/dt= -\nu{\cal F}^{(0)}/(M\partial E^{(0)}/\partial\omega)$, that follows from Eq.~\eqref{Omega dot}. An $n$PN term is defined analogously. On the other hand, an $n$th-order self-force term ($n$SF) is computed from metric perturbations in the binary's spacetime that are of \emph{absolute} order $\nu^n$.

\begin{figure*}[!bth]
{\centering
\includegraphics[width=.98\textwidth,trim={0 20pt 0 0}]{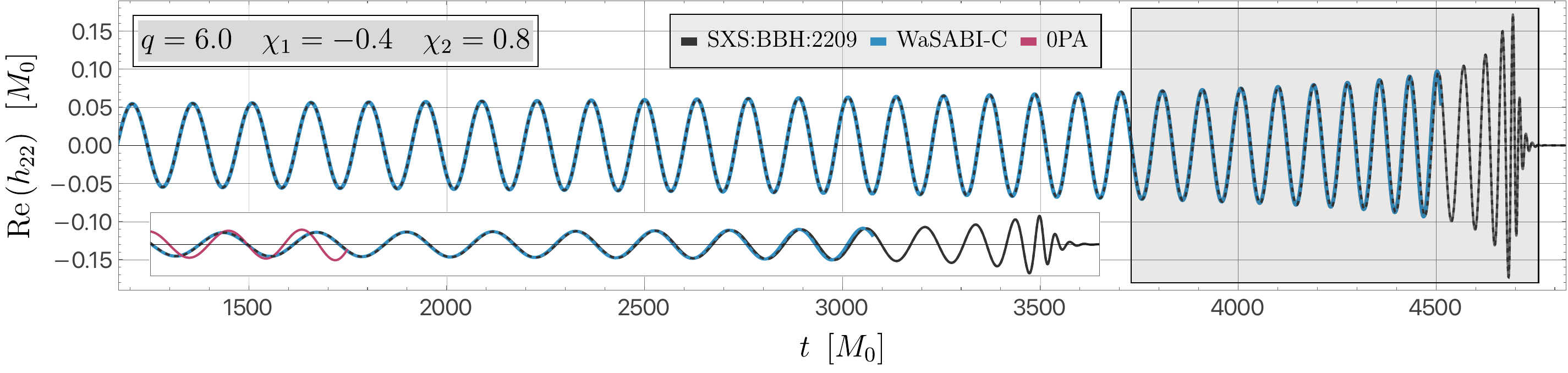}}
\caption{$(2,2)$ mode of the gravitational waveform, with the NR simulation SXS:BBH:2209 in black and the \texttt{WaSABI-C} model in blue. The inset zooms on the gray shaded region, with an adiabatic 0PA model in purple for reference. The waveforms have been aligned at the reference time of the NR simulation using the same alignment procedure as described in Ref.~\cite{PaperIII}.
}\label{fig:WF}
\end{figure*}

This hybridization procedure is set entirely with gauge-invariant functions of invariant variables, and it preserves as much as possible of the structure of the \emph{exact} binary evolution equations such as Eq.~\eqref{Omega dot}, which only rely on exact laws of general relativity. In particular, we stress that we do not directly expand the right-hand side of the differential equations~\eqref{eq:ODE}. Doing so would incur a loss of accuracy when expanding $F_\omega(J_a)$ in either the PN or SF limit. $F_\omega(J_a)\approx M d\Omega/dt$ has effectively the same structure as Eq.~\eqref{Omega dot}, which encounters a divergence at the innermost stable circular orbit, where $\partial E/\partial \omega=0$; SF and PN expansions of the fraction spoil this pole structure. To circumvent this, we build composite expansions of the SF and PN series for the binding energy and fluxes and then derive Eq.~\eqref{eq:ODE} \emph{without} re-expanding the fraction in an SF or PN series. Reference~\cite{Burke:2023lno} and the companion paper~\cite{PaperII} show the significant accuracy enhancement this yields.

When designing our hybridization scheme, we also tested different expansions in which we held alternative spin variables fixed, such as $\tilde{\chi}_i\equiv (m_i/M) \chi_i$, $s\equiv (m_1/M) \tilde{\chi}_1+(m_2/M) \tilde{\chi}_2$, and $\sigma\equiv \tilde \chi_2 -\tilde \chi_1$. We found no significant differences in our waveforms, contrasting with notable improvements seen in Ref.~\cite{PaperII} when working with more limited input data.

\emph{SF and PN data.} Modeling EMRIs and other asymmetric-mass systems requires 1PA accuracy in the SF expansion~\cite{LISAConsortiumWaveformWorkingGroup:2023arg,Burke:2023lno}, which involves 1SF dissipative effects (0PA), 1SF conservative and 2SF dissipative effects (1PA), and both conservative and dissipative  linear-in-$\chi_2$ (but \emph{exact} in $\chi_1$) effects (1PA)~\cite{Mathews:2021rod}. For the quasicircular, (anti)aligned-spin binaries we consider, these ingredients reduce to the 1SF binding energy, 1SF and 2SF fluxes, and linear-in-$\chi_2$ contributions to the binding energy and fluxes. All of these are readily computed~\cite{Barack:2018yvs,TeukolskyPackage,Nasipak:pybhpt,Taracchini:2014zpa,Nasipak:2023kuf,Le_Tiec_2012,Shah:2012gu,Isoyama:2014mja,vandeMeent:2015lxa,Piovano:2021iwv} \emph{except} the 2SF fluxes, which are only known in the case of a nonspinning primary BH ($\chi_1=0$)~\cite{Warburton:2021kwk}. An essential goal of our model is to provide an accurate substitute for that missing 2SF flux for finite $\chi_1$. Similarly, the waveform amplitudes are only known to 2SF order in the case $\chi_1=0$.

On the PN side, the binding energy and fluxes are known to 4PN in the spinning sector and 4.5PN in the nonspinning sector~\cite{blanchet2024postnewtoniantheorygravitationalwaves,PNpedia,Blanchet:2023bwj,Blanchet:2023sbv,Marsat:2013caa,Blanchet:2006gy,Bohe:2013cla,Cho:2022syn,Bohe:2015ana,Cho:2021mqw,Cho:2022syn,Tagoshi:1997jy,Alvi:2001mx,Porto:2007qi,Chatziioannou:2012gq,Saketh:2022xjb,Damour:2014jta,Marchand:2017pir,Foffa:2019yfl,Cho:2022syn,Bohe:2013cla,Boh__2013,Bohe:2015ana,Marsat:2014xea} while the amplitudes are known to 3.5PN \cite{Henry:2022ccf} except for the 22 mode known to 4PN \cite{Henry:2022ccf,Blanchet:2023bwj,Blanchet:2023sbv,Warburton:2024xnr}. Completing the 1PA dynamics would not require using all this PN data, but 1PA models lose accuracy for long signals that extend into the weak-field regime due to omitted PN information~\cite{Albertini:2022rfe}. Such signals are important for moderate and intermediate mass ratios~\cite{Chapman-Bird:2025xtd}. 

Hence, to model the broadest range of systems, we use all the above information when building our composite expansions. This includes all available PN terms (excluding quartic-in-spin terms~\cite{Siemonsen:2022fsa}) as well as 2SF flux results for $\chi_1=0$ and linear-in-$\chi_2$ results for generic $\chi_1$, which we obtain by employing the codes developed in Refs.~\cite{vandeMeent:2015lxa,Warburton:2021kwk,Piovano:2024yks}. We provide more specific details of the data used in \texttt{WaSABI-C} in the companion papers~\cite{PaperII,PaperIII} and the Supplementary Material. 

\emph{Results.}   To assess our model's accuracy, we compare against numerical relativity (NR) simulations from the SXS catalog~\cite{SXSPackage_v2025.0.18,SXSCatalogPaper_3} with IDs detailed in Table~\ref{tab:sxscontent} of the Supplementary Material.

In Fig.~\ref{fig:WF}, we plot the real part of the $(2,2)$ mode of our model waveform (in units of the initial total mass $M_0$), against the NR simulation SXS:BBH:2515 with large mass ratio $q=6$, primary spin $\chi_1=-0.4$ and secondary spin $\chi_2=0.8$. The waveforms have been aligned at reference time by performing the least-square error procedure on the waveform frequency as described in one of our companion papers~\cite{PaperIII}. In \texttt{WaSABI-C}, the waveform is ended by default at the maximum of $F_\omega$. Qualitatively, we see a very good agreement of both the waveform phase and the waveform amplitude. We also show for reference the real part of the $(2,2)$-mode of an adiabatic leading-order self-force model (labeled `0PA'), which dephases much more rapidly than \texttt{WaSABI-C} over the course of the inspiral.

We use the mismatch of the (2,2) mode as our main metric for measuring the faithfulness of \texttt{WaSABI-C} across the parameter space. The mismatch between two waveforms $X$ and $Y$ is given by
\begin{equation}\label{eq:mismatch}
	\mathcal{M}^\text{insp}_{22}=1-\max_{\Delta t,\Delta\varphi}\frac{\langle h^X_{22}({\Delta t,\Delta \varphi}),h_{22}^Y\rangle}{\| h_{22}^X({\Delta t,\Delta\varphi})\|\| h_{22}^Y\|},
\end{equation}
which is optimized over time and phase shifts $\Delta t$ and $\Delta\varphi$. The scalar product is given in Fourier domain by
\begin{equation}
	\langle h_{22}^X,h_{22}^{Y}\rangle=\frac{2}{\pi} \text{Re}\int_{\omega_i}^{\omega_f}  \frac{\left({\mathcal{F}[h_{22}^X]}(\omega)\right)^* {\mathcal{F}[h_{22}^{Y}]}(\omega)}{S_n(\omega/(2\pi))}d\omega,
\end{equation}
where $\mathcal{F}[h_{22}^X]$ is the Fourier transform of $h_{22}^X$ and $S_n$ is the power spectral density.

\begin{figure}[tb]
 \center
 \includegraphics[width=0.48\textwidth,trim={0 20pt 0 0},clip]{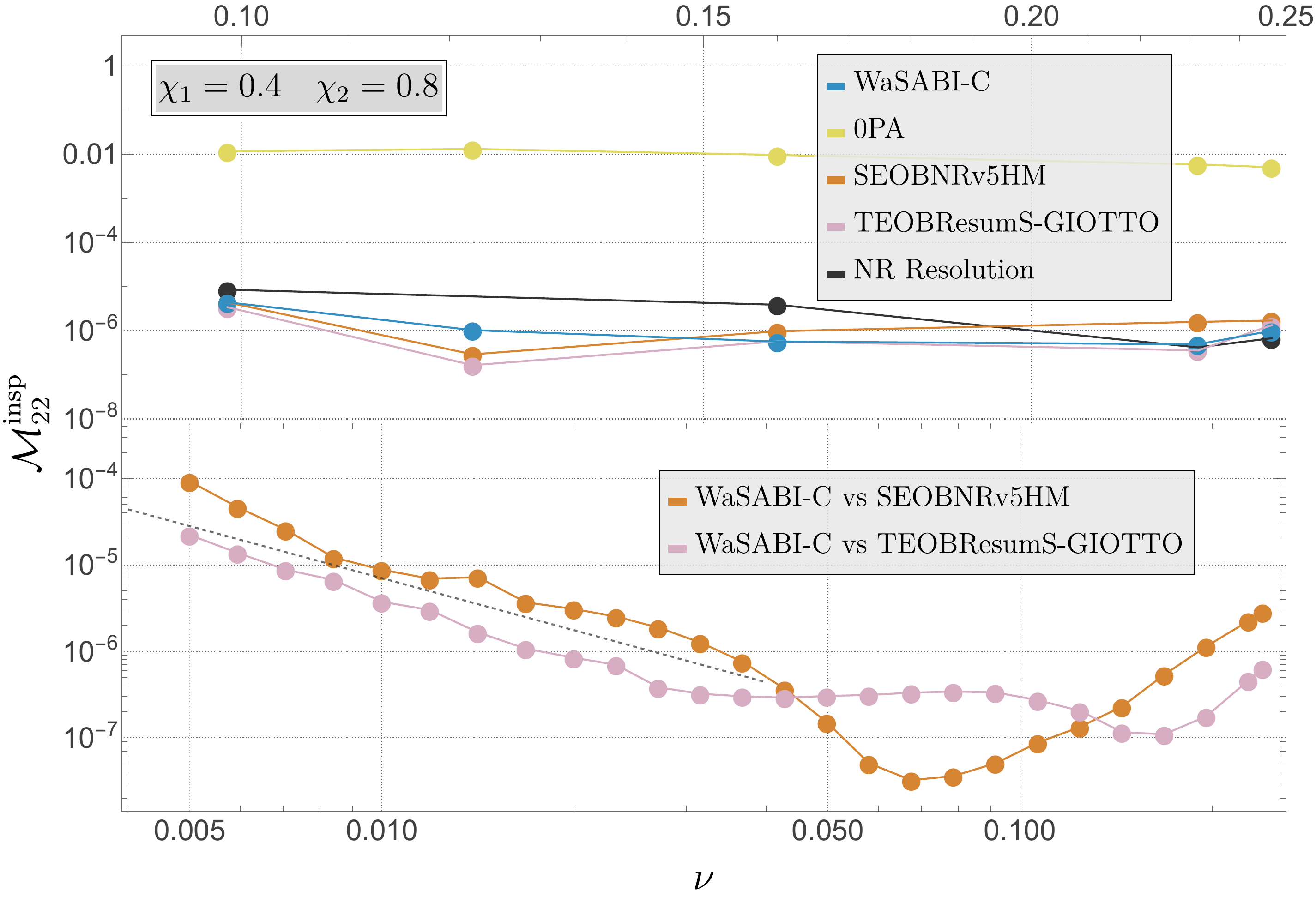}
 \caption{Mismatch as a function of mass ratio on a fixed waveform frequency range $[0.044/M,0.12/M]$ for inspirals with initial spins $\chi_{1,0}=0.4$, $\chi_{2,0}=0.8$, using a flat power spectral density. The frequency range corresponds to an orbital separation going from $12.7M$ to $6.3M$. Upper panel: mismatches of the four models \texttt{WaSABI-C}, 0PA, \texttt{SEOBNRv5HM} and \texttt{TEOBResumS-GIOTTO} against 5 NR simulations from the SXS catalog. 
 The mismatch between the two highest NR resolutions is indicated for reference. Lower panel: mismatches between \texttt{WaSABI-C} and either \texttt{SEOBNRv5HM} or \texttt{TEOBResumS-GIOTTO}, which extend to lower mass ratios than covered by NR. The dotted line is a reference power law~$\nu^{-2}$.}\label{fig:mismatchvsmassratio}
 
\end{figure}

In Fig. \ref{fig:mismatchvsmassratio}, we plot the mismatch $\mathcal{M}^\text{insp}_{22}$ against NR templates of four distinct models as a function of the symmetric mass ratio $\nu$ for a specific spin configuration. 
The inspiral waveforms produced by \texttt{WaSABI-C} are  dramatically more accurate than the 0PA model as the mismatches differ by approximately four orders of magnitude. In the comparable-mass range ($0.1<\nu<0.25$), \texttt{WaSABI-C} produces very similar inspiral waveforms as other state-of-the-art semi-analytical models such as \texttt{SEOBNRv5HM}~\cite{Pompili:2023tna} and \texttt{TEOBResumS-GIOTTO}~\cite{Nagar:2023zxh,Riemenschneider:2021ppj} as the mismatches are comparable. In contrast, below  $\nu\lesssim 0.05$ the mismatch between \texttt{WaSABI-C} and both EOB models scales as $1/ \nu^2$. Since neither \texttt{SEOBNRv5HM} nor \texttt{TEOBResumS-GIOTTO} uses full 0PA fluxes in their dynamics, they both differ from \texttt{WaSABI-C} at leading, adiabatic order in the small-$\nu$ limit. This leads to an accumulated phase error which grows like $1/\nu$~\cite{Hinderer:2008dm}, and hence a mismatch against \texttt{WaSABI-C} that grows like $1/\nu^2$~\cite{Mitman:2025tmj}. This is clear evidence that \texttt{WaSABI-C} is more accurate than both EOB models below $\nu\lesssim0.05$.

\begin{figure}[tb]
 \center
 \includegraphics[width=0.45\textwidth,trim={0 -20pt 0 0}]{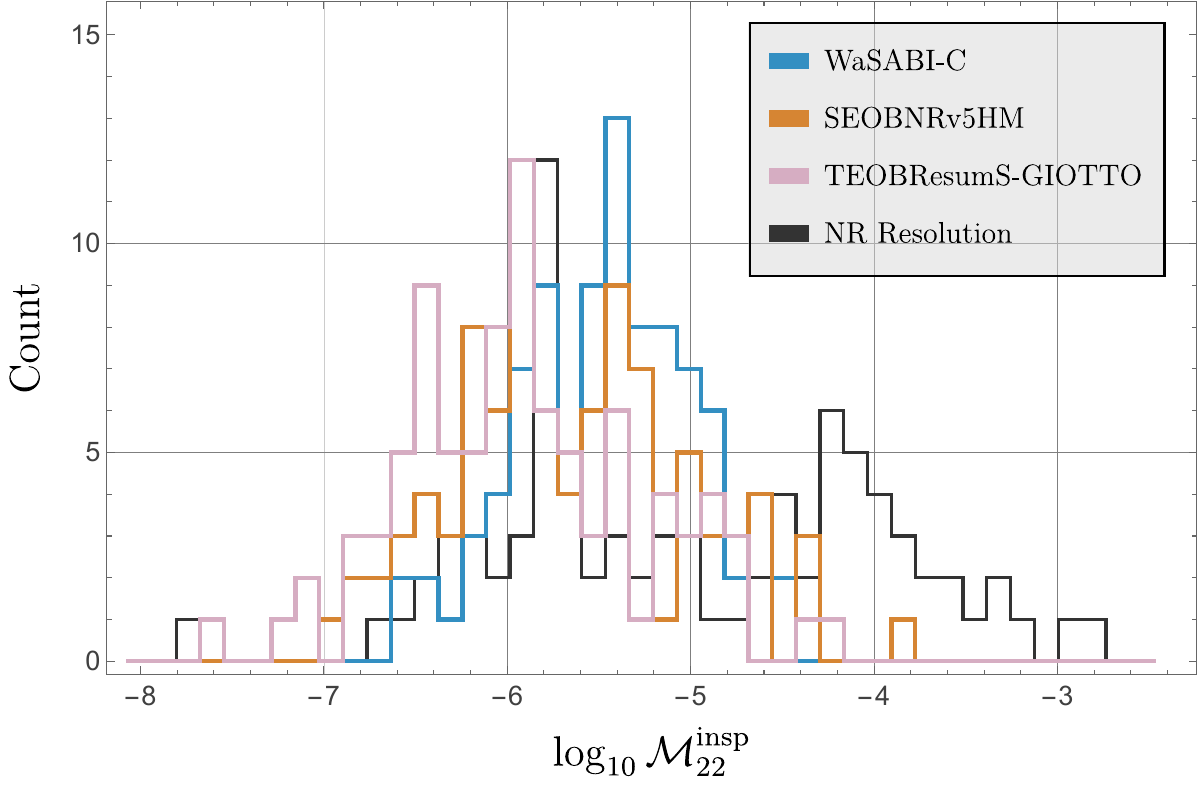}
  \includegraphics[width=0.45\textwidth,trim={0 10pt 0 0}]{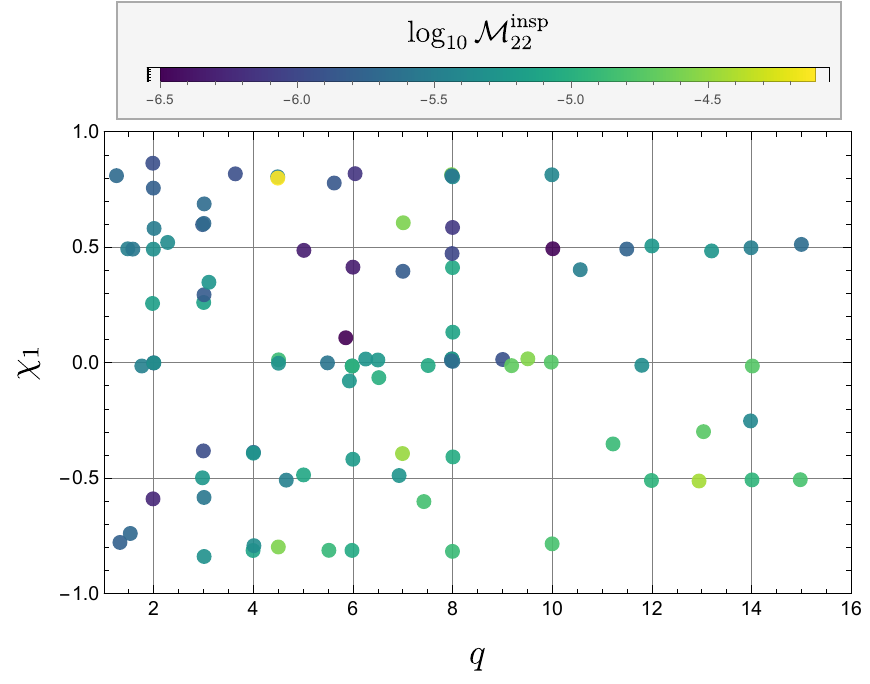}

 \caption{Upper panel: histogram of mismatches between NR simulations and either \texttt{WaSABI-C}, \texttt{SEOBNRv5HM} or \texttt{TEOBResumS-GIOTTO}. For each simulation, the frequency window ranges from the reference frequency of the simulation to the breakdown frequency of \texttt{WaSABI-C}. We used the Virgo O5 design curve~\cite{LVK_PSD} and set the initial total mass of the system at $M_0=20 M_\odot$. Bottom panel: Population of NR simulations used for this study. The secondary spin axis dimension has been suppressed for clarity.  
 For each point on parameter space, we report the mismatch between \texttt{WaSABI-C} and the corresponding SXS simulation. All simulation IDs can be found in the supplementary material.}\label{fig:histogram} 
\end{figure}

In Fig. \ref{fig:histogram}, we benchmark \texttt{WaSABI-C} across the parameter space covered by NR simulations. We selected 93 SXS simulations covering as uniformly as possible the parameter space $(q,\chi_1,\chi_2)$ and computed the mismatch of \texttt{WaSABI-C} against those templates. 
In the comparable-mass range, \texttt{WaSABI-C} performs slightly worse than both EOB models, with a median mismatch of $4.5\times10^{-6}$, as compared to $1.3\times 10^{-6}$ and $2.0\times 10^{-6}$ for \texttt{TEOBResumS-GIOTTO} and \texttt{SEOBNRv5HM}, respectively. 
We do not identify any particular region in parameter space where the mismatch is significantly higher except for a slight trend towards retrograde orbits, as already found in Ref. \cite{PaperIII}. The largest recorded mismatch against NR peaks at $7.7\times10^{-5}$ for $\chi_1=0.8$, $\chi_2=0$, $q=4.5$. 

From these analyses, we find \texttt{WaSABI-C} to be competitive with current publicly available EOB models for comparable masses but expect it to outperform those models for mass ratios above $q\gtrsim 20$. We expect the accuracy to be highest when the primary black hole is nonspinning, as our model  includes second-order SF energy flux for non-spinning binaries~\cite{Warburton:2021kwk}. We point the reader to our accompanying paper~\cite{PaperIII} for an analysis of the expected impact of second-order SF fluxes across all primary spins.

\emph{Discussion}. 
We have introduced a new gravitational waveform model for spin-aligned, quasicircular inspirals of asymmetric BH binaries, built from a hybridization of the latest first- and second-order SF results with high-order PN expansions. By combining SF data with PN information, our model accurately captures the dynamics of binaries across a wide range of mass ratios, from the extreme-mass-ratio regime to intermediate mass ratios and even comparable masses, making it relevant for a range of sources for future gravitational-wave observatories such as LISA and third-generation ground-based detectors. Our model can (by design) be readily incorporated into the FEW package, enabling generation of LISA-length signals in tens of milliseconds.

One of the key achievements of our model is its strong performance when benchmarked against state-of-the-art NR  simulations. Across the mass-ratio parameter space, we find that the waveform mismatches with NR are comparable to those obtained with \texttt{SEOBNRv5HM} and \texttt{TEOBResumS-GIOTTO}, two of the most accurate EOB models currently available (which include some calibration to NR in their inspiral dynamics, though we note EOB's calibration to NR is mostly confined to the merger-ringdown phase of the waveform). This level of agreement underscores the efficacy of our SF/PN hybrid approach in capturing the essential physics of spin-aligned inspirals. Our method is conceptually simple, involves no NR calibration, and does not require the often intricate resummations of the waveform employed in EOB models, demonstrating the power of employing exact SF results.

There are clear paths to future improvements. First, our treatment of the primary BH's spin is limited to first-order SF results. Second-order results will be required to assess---and potentially provide critical enhancements to---our model's accuracy in the EMRI regime for rapidly spinning primaries~\cite{PaperIII}.

Second, our models are restricted to spin-aligned (i.e., non-precessing), quasicircular orbits. While this simplifies the problem and is a reasonable approximation for certain astrophysical scenarios, realistic inspirals---especially those formed through dynamical capture in dense stellar environments~\cite{LISA:2022yao}---may exhibit significant eccentricity and spin-induced precession, and certain classes of EMRIs can have eccentricities close to unity~\cite{Qunbar:2023vys,Mancieri:2024sfy,Mancieri:2025cmx}. Extending the models to incorporate eccentricity~\cite{Mathews2025eccentric} and misaligned spins~\cite{Piovano:2025aro,Piovano:2024yks} will be essential for maximizing their applicability to a broader range of GW sources.

Finally, our models are restricted to the inspiral stage of the binary evolution. For intermediate-mass-ratio and comparable-mass systems, the final merger and ringdown can represent a significant fraction of the binary's observable signal, particularly for ground-based detectors. Our models must be extended to include these final stages (e.g., using methods from~\cite{Rifat:2019ltp,Islam:2022laz,Rink:2024swg,Kuchler:2024esj,Honet:2025dho,Kuchler:2025hwx,Roy:2025kra,Paul:2024ujx,Iglesias:2025tnt,Islam:2025tjj}).

Our public Mathematica package \href{https://bhptoolkit.org/WaSABI/}{\textsc{WaSABI}}~\cite{BHPT_WaSABI} provides a modular and extensible platform for incorporating these improvements.

In summary, our models represent a significant step forward in the construction of high-accuracy inspiral waveforms rooted in SF theory. They bridge the gap between perturbative and numerical relativity methods and sets the stage for further advances in modeling the rich dynamics of asymmetric, spinning compact binaries.

\emph{Acknowledgments.} We thank C. Kavanagh, D. Trestini and Z. Nasipak for insightful discussions. 
MvdM acknowledges financial support by 
the VILLUM Foundation (grant no. VIL37766),
the DNRF Chair program (grant no. DNRF162) by the Danish National Research Foundation,
and the EU
’s Horizon ERC Synergy Grant “Making Sense of the Unexpected in the Gravitational-Wave Sky” grant GWSky–101167314.  
The Center of Gravity is a Center of Excellence funded by the Danish National Research Foundation under grant No. 184.
AP acknowledges the support of a Royal Society University Research Fellowship and the ERC Consolidator/UKRI Frontier Research Grant GWModels (selected by the ERC and funded by UKRI [grant EP/Y008251/1]). 
LH acknowledges the support of the FNRS through a FRIA doctoral grant. G.C. is Research Director of the FNRS.
G.C. and G.A.P. acknowledge the support of the Win4Project grant ETLOG of the Walloon Region for the Einstein Telescope.
JM acknowledges support from the NUS Faculty of Science, under the grant 22-5478-A0001.
This work makes use of the Black Hole Perturbation Toolkit~\cite{BHPToolkit} and PNpedia~\cite{PNpedia}. We made use of the Tycho supercomputer hosted at the SCIENCE HPC center (U. Copenhagen) as well as the Lyra cluster (ULB).

\bibliography{ThisBib}

\clearpage

\appendix

\onecolumngrid

\section*{Supplementary Material}
\label{app:content}
\setcounter{page}{1}
\setcounter{equation}{0}

In this Supplementary Material, we provide (i) more details of our hybridization scheme (summarizing the method from the companion paper~\cite{PaperIII}), (ii) more details of which SF and PN data we utilize, and (iii) the list of SXS simulations we use in our accuracy tests.

\vspace{-0.3cm}
\subsection*{Hybridization}

When building composite expansions for the energy, fluxes, and amplitudes, we first adimensionalize all quantities by factoring out an appropriate power of the total mass $M$. In place of $\omega$, we use $x \equiv \omega^{2/3}$ and write any such gauge-invariant dimensionless quantity as $Q(\nu,x,\chi_i)$. The SF and PN approximations of such a quantity correspond to small-$\nu$ and small-$x$ expansions, and a combined SF$\vert$PN approximation is an expansion in both $\nu$ and $x$:
\begin{subequations}
\begin{align}
Q^\text{SF}_{\tiny \!\!\begin{array}{c} \bar k \, \overline{l}_1\, \overline{l}_2 \vspace{-1.5pt}\\ \underline{k} \mbox{\hspace{1.5pt}}  \,  \underline{l}_1\, \underline{l}_2\end{array}}\! &=\nu^K  \sum_{l_i=\underline{l}_i}^{\overline{l}_i}\sum_{k=\underline{k}}^{\bar k}  Q^{(k)}_{l_1l_2}(x)\nu^k \chi_1^{l_1}\chi_2^{l_2} ,\\
Q^\text{PN}_{\tiny \!\!\begin{array}{c} \bar n \, \overline{l}_1\, \overline{l}_2 \vspace{-1.5pt}\\ \underline{n} \mbox{\hspace{1.5pt}}  \,  \underline{l}_1\, \underline{l}_2\end{array}}\! &= x^{N/2}\sum_{l_i=\underline{l}_i}^{\overline{l}_i} \sum_{n=\underline{n}}^{\bar n}  Q^{\frac{n}{2}\text{PN}}_{l_1l_2}(\log x) x^{n/2}\chi_1^{l_1}\chi_2^{l_2} ,\\
Q^\text{SF$\vert$PN}_{\tiny \!\!\begin{array}{c} \bar k  \, \bar n \, \overline{l}_1\, \overline{l}_2 \vspace{-1.5pt}\\ \underline{k} \, \underline{n} \mbox{\hspace{1.5pt}}  \,  \underline{l}_1\, \underline{l}_2\end{array}}\! &=\nu^K x^{N/2}\sum_{l_i=\underline{l}_i}^{\overline{l}_i}  \sum_{k=\underline{k}}^{\bar k}\sum_{n=\underline{n}}^{\bar n} Q^{k \vert n}_{l_1l_2}(\log x)\nu^k x^{n/2}\chi_1^{l_1}\chi_2^{l_2} 
\end{align}
\end{subequations}
where $K$ and $N/2$ denote the leading power in the SF and PN series, respectively; underlined indices denote the lowest terms included; and overlined indices denote the highest terms included. We define the hybridized quantity $Q^\text{SF$+$PN}$ as the unique sum of the SF and PN approximations 
minus their common terms, 
\begin{align}\label{QSFPN}
\!Q^\text{SF$+$PN}=\!\!\!\!\!\!\sum_{\underline{k},\underline{n},\underline{l}_i,\overline{k},\overline{n},\overline{l}_i}\left( Q^\text{SF}_{\tiny \!\!\begin{array}{c} \bar k  \,  \overline{l}_1\, \overline{l}_2 \vspace{-1.5pt}\\ \underline{k} \mbox{\hspace{1.5pt}}  \,  \underline{l}_1\, \underline{l}_2\end{array}}\!+Q^\text{PN}_{\tiny \!\!\begin{array}{c} \bar n \, \overline{l}_1\, \overline{l}_2 \vspace{-1.5pt}\\ \underline{n} \mbox{\hspace{1.5pt}}  \,  \underline{l}_1\, \underline{l}_2\end{array}}\! - Q^\text{SF$\vert$PN}_{\tiny \!\!\begin{array}{c} \bar k  \, \bar n \, \overline{l}_1\, \overline{l}_2 \vspace{-1.5pt}\\ \underline{k} \, \underline{n} \mbox{\hspace{1.5pt}}  \,  \underline{l}_1\, \underline{l}_2\end{array}}\!   \right).  
\end{align}

In most cases, $\overline{k}$ denotes a $\overline{k}$PA order expansion, and $\overline{n}$ denotes an $\frac{\overline{n}}{2}$PN expansion. But in some cases, such as the fluxes through the horizon, PN orders instead count from the leading power of $x$ in the total energy flux, rather than from the leading power of $x$ in the specific horizon flux.

\vspace{-0.3cm}
\subsection*{SF and PN data}

\begin{figure}[!htb]
{\centering
\includegraphics[width=.95\textwidth]{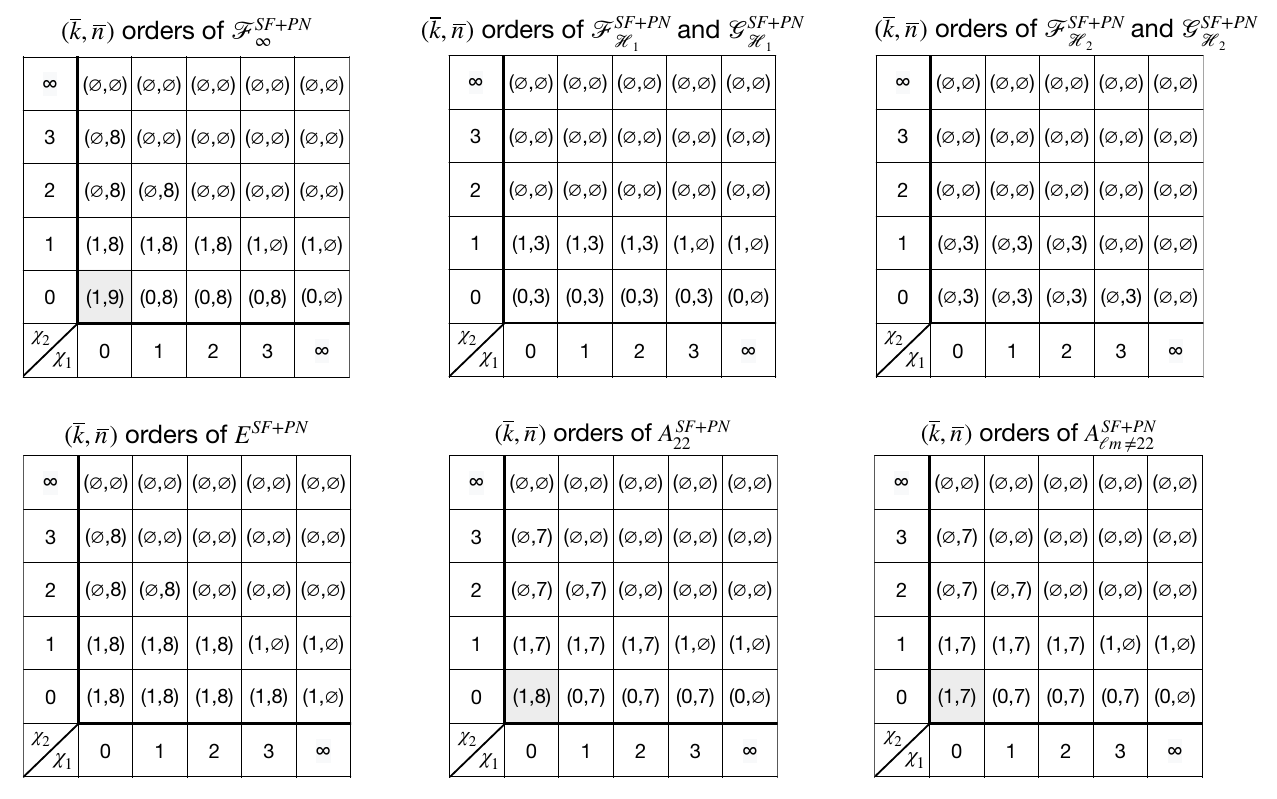}}
\caption{Content of the hybrid functions $Q=\{\mathcal F_\infty , \mathcal F_{\mathcal H_i} , \mathcal G_{\mathcal H_i},E, A_{\ell m}\}$ built as part of the \texttt{WaSABI-C} model. The information is stored as follows. Each entry $(\overline{k},\overline{n})$ for each pair $(\chi_1,\chi_2)$ in a given table denotes the maximal orders $(\overline k,\overline n)$ being summed over in the hybridization of the corresponding function \eqref{QSFPN}. For quantities other than ${\cal F}_{{\cal H}_i}$ and ${\cal G}_{{\cal H}_i}$, $(\overline{k},\overline{n})$ denotes $\overline{k}$PA and $\frac{\overline{n}}{2}$PN information. As the spin content $\chi_i$, $i=1,2$, is limited to cubic order in the current PN approximation, we denote as $\infty$ all powers of spin larger than $3$ up to $\infty$. The emptyset symbol $\emptyset$ means that no data is being used. Shaded gray entries indicate sectors where 2SF information is being used.
}\label{WaSABI-C_Content}
\end{figure}

We use the highest-order available SF and PN data for each quantity, which implies the following bounds on the above approximations: PN data is limited to $4$PN order ($\overline n =8$), except for the energy flux at infinity given at $4.5$PN order ($\overline n =9$). This limits the spin interactions to SSSS interactions ($\overline{l}_1+\overline{l}_2 \leq 4$), but we neglect the latter (which were computed in Ref.~\cite{Siemonsen:2022fsa}) and restrict ourselves to the SO, SS, and SSS interactions; first-order self-force data is valid for an arbitrary primary spin ($\overline l_1 = \infty$); first- and second-order self-force data are limited to linear order in the secondary spin ($\overline l_2=1$). This is summarized in Fig.~\ref{WaSABI-C_Content}. Note that the spin-induced multipole moments of both black holes are included as part of the PN expansion to the truncated PN order considered. In the SF expansion, such spin-induced multipoles are automatically included for the primary spin at 0PA order (since the secondary evolves around Kerr) while they are not included for the secondary. 

Most of the SF and PN expressions used as inputs are detailed in the Appendices of \cite{PaperIII}. With respect to the $\texttt{WaSABI-C}$ (v0.9) model of \cite{PaperIII}, the $\texttt{WaSABI-C}$ (v1.0) model in this Letter contains the following four additional ingredients: 
\begin{enumerate}
\item the PN expressions for the energy flux at infinity \cite{Blanchet:2006gy,Bohe:2013cla,Marsat:2013caa,Blanchet:2013haa,Marsat:2014xea,Marchand:2016vox,Marchand:2020fpt,Larrouturou:2021dma,Larrouturou:2021gqo,Cho:2022syn,Blanchet:2023bwj,Blanchet:2023sbv,Trestini:2023wwg,Warburton:2024xnr,Cunningham:2024dog}, the binding energy  \cite{Bernard:2017ktp,Damour:2014jta,Marsat:2014xea,Cho:2022syn,Trestini:2025nzr}, and the amplitudes \cite{Favata:2008yd,Blanchet:2008je,Faye:2012we,Faye:2014fra,Henry:2021cek,Henry:2022dzx,Henry:2022ccf,Blanchet:2023bwj,Blanchet:2023sbv} now include all known secondary spin contributions; 
\item the mode amplitudes, 1SF energy fluxes at infinity and at the horizon and redshift data are calculated on refined grids using the code of~\cite{Piovano:2024yks,repoHJproject}
and the metric reconstruction code of~\cite{vandeMeent:2015lxa}. The data was generated and interpolated on a 36 by 36 grid of Chebyshev nodes in $\log(1-a)$ and $(r_{\rm isco}/r)^{1/2}$, leading to an estimated relative interpolation error $\lesssim 10^{-10}$, $\lesssim 10^{-10}$ and $\lesssim 10^{-7}$ for amplitudes, fluxes and redshift, respectively;
\item the analytic linear-in-secondary-spin (1PA) contributions to the SF binding energy were added. Moreover, we incorporate the linear-in-secondary-spin (1PA) contribution to the SF energy fluxes at infinity and at the horizon as well as to the amplitudes, which are computed using the code of~\cite{Piovano:2024yks,repoHJproject}. The data was generated and interpolated on the same Chebyshev grid adopted for the redshift data. The estimated relative interpolation error for the  linear-in-secondary-spin (1PA) contribution to the SF energy fluxes is $\lesssim 10^{-7}$; 
\item finally the 2SF energy flux at infinity and the 2SF amplitudes for non-spinning binaries \cite{Warburton:2021kwk,Wardell:2021fyy} were added to the model. 
\end{enumerate}

\clearpage

\subsection*{SXS simulations}
\vspace{-0.4cm}
\begin{table*}[!hb]
\begin{ruledtabular}
\begin{tabular}{cccccc} 
{SXS:BBH:2670} & {SXS:BBH:2677} & {SXS:BBH:3136} &
{SXS:BBH:2480} & {SXS:BBH:2474} & {SXS:BBH:1438} \\
{SXS:BBH:0189} & {SXS:BBH:1436} & {SXS:BBH:2475} &
{SXS:BBH:1452} & {SXS:BBH:2486} & {SXS:BBH:2472} \\
{SXS:BBH:1464} & {SXS:BBH:2018} & {SXS:BBH:2488} &
{SXS:BBH:1437} & {SXS:BBH:0371} & {SXS:BBH:0203} \\
{SXS:BBH:2757} & {SXS:BBH:2155} & {SXS:BBH:3622} &
{SXS:BBH:4260} & {SXS:BBH:0202} & {SXS:BBH:4235} \\
{SXS:BBH:4430} & {SXS:BBH:2508} & {SXS:BBH:2515} &
{SXS:BBH:3630} & {SXS:BBH:3891} & {SXS:BBH:3865} \\
{SXS:BBH:2141} & {SXS:BBH:2179} & {SXS:BBH:1962} &
{SXS:BBH:2786} & {SXS:BBH:2495} & {SXS:BBH:1441} \\
{SXS:BBH:0331} & {SXS:BBH:2466} & {SXS:BBH:1448} &
{SXS:BBH:2696} & {SXS:BBH:1470} & {SXS:BBH:2469} \\
{SXS:BBH:1485} & {SXS:BBH:2467} & {SXS:BBH:1440} &
{SXS:BBH:2209} & {SXS:BBH:3122} & {SXS:BBH:2494} \\
{SXS:BBH:0525} & {SXS:BBH:0513} & {SXS:BBH:2479} &
{SXS:BBH:2642} & {SXS:BBH:2470} & {SXS:BBH:2484} \\
{SXS:BBH:1468} & {SXS:BBH:0185} & {SXS:BBH:2476} &
{SXS:BBH:2482} & {SXS:BBH:1460} & {SXS:BBH:2478} \\
{SXS:BBH:2160} & {SXS:BBH:4432} & {SXS:BBH:2134} &
{SXS:BBH:2502} & {SXS:BBH:2161} & {SXS:BBH:2168} \\
{SXS:BBH:2701} & {SXS:BBH:2755} & {SXS:BBH:0206} &
{SXS:BBH:4236} & {SXS:BBH:2132} & {SXS:BBH:2569} \\
{SXS:BBH:2110} & {SXS:BBH:2119} & {SXS:BBH:2186} &
{SXS:BBH:1152} & {SXS:BBH:2127} & {SXS:BBH:1932} \\
{SXS:BBH:0612} & {SXS:BBH:1961} & {SXS:BBH:0615} &
{SXS:BBH:4284} & {SXS:BBH:3924} & {SXS:BBH:2668} \\
{SXS:BBH:3127} & {SXS:BBH:1445} & {SXS:BBH:1427} &
{SXS:BBH:3128} & {SXS:BBH:1444} & {SXS:BBH:2464} \\
{SXS:BBH:2490} & {SXS:BBH:1428} & {SXS:BBH:1426} \\
\end{tabular}
\end{ruledtabular}
\caption{List of SXS simulation IDs used in this work. All simulations are available from the third SXS catalog \cite{SXSPackage_v2025.0.18,SXSCatalogPaper_3} and were released in Refs. \cite{Varma_2019,Varma_2018,Blackman_2017,Varma_2019,Boyle_2019,Blackman_2015,Kumar_2015,Yoo_2022}.}\label{tab:sxscontent}
\end{table*}

\end{document}